\documentclass[journal]{IEEEtran}

\usepackage[utf8]{inputenc}
\usepackage{color}
\usepackage{xcolor}
\usepackage{array}
\usepackage{verbatim}
\usepackage{float}
\usepackage{amsmath}
\usepackage{amsthm}
\usepackage{amssymb}
\usepackage{graphicx}
\usepackage{longtable}
\usepackage{multirow}
\usepackage{booktabs}
\usepackage{longtable}

\usepackage[unicode=true,
bookmarks=false,
breaklinks=false,pdfborder={0 0 1},colorlinks=false]
{hyperref}
\hypersetup{
	colorlinks,bookmarksopen,bookmarksnumbered,citecolor=blue,urlcolor=blue}
\usepackage{cite}

\usepackage{lipsum}
\usepackage{mathtools}
\usepackage{cuted}

\usepackage{algorithmic}
\usepackage{longtable}

\floatstyle{ruled}
\newfloat{algorithm}{tbp}{loa}
\providecommand{\algorithmname}{Algorithm}
\floatname{algorithm}{\protect\algorithmname}

\makeatletter
\let\oldforeign@language\foreign@language
\DeclareRobustCommand{\foreign@language}[1]{%
	\lowercase{\oldforeign@language{#1}}}

\let\oldforeign@language\foreign@language
\DeclareRobustCommand{\foreign@language}[1]{%
	\lowercase{\oldforeign@language{#1}}}

\ifCLASSINFOpdf
\else
\fi

\@ifundefined{showcaptionsetup}{}{%
	\PassOptionsToPackage{caption=false}{subfig}}
\usepackage{subfig}

\usepackage{balance}

\ifCLASSINFOpdf
\else
\fi

\ifCLASSINFOpdf
\else
\fi

\def\ps@IEEEtitlepagestyle{%
	\def\@oddhead{\parbox[t][\height][t]{\textwidth}{\centering \scriptsize
			Personal use of this material is permitted. Permission from the author(s) and/or copyright holder(s), must be obtained for all other uses. Please contact us and provide details if you believe this document breaches copyrights.\\
			\noindent\makebox[\linewidth]{}
		}\hfil\hbox{}}%
	\def\@evenhead{\scriptsize\thepage \hfil \leftmark\mbox{}}%
	\def\@oddfoot{\parbox[t][\height][l]{\textwidth}{
			\vspace{-20pt}{\rule{\textwidth}{0.4pt}}\\ \footnotesize{\bf{\footnotesize\textcolor{red}{L. Tirel, A. M. Ali, and H. A. Hashim, "Novel Hybrid Integrated Pix2Pix and WGAN Model with Gradient Penalty for Binary Images Denoising," Systems and Soft Computing, vol. 6, pp. 200122, 2024.}}} doi: \href{https://doi.org/10.1016/j.sasc.2024.200122}{10.1016/j.sasc.2024.200122}\\\\
			\noindent\makebox[\linewidth]
		}\hfil\hbox{}}%
	\def\@evenfoot{\MYfooter}}

\makeatother
\begin{document}
	\bstctlcite{IEEEexample:BSTcontrol}

\title{Novel Hybrid Integrated Pix2Pix and WGAN Model with Gradient Penalty for Binary Images Denoising}

\author{Luca Tirel, Ali Mohamed Ali, and Hashim A. Hashim
	\thanks{This work was supported in part by the National Sciences and Engineering Research Council of Canada (NSERC), under the grants RGPIN-2022-04937.}
	\thanks{L. Tirel, A. M. Ali, and H. A. Hashim are with the Department of Mechanical
		and Aerospace Engineering, Carleton University, Ottawa, ON, K1S-5B6,
		Canada (e-mail: hhashim@carleton.ca).}
}



\maketitle
\begin{abstract}
This paper introduces a novel approach to image denoising that leverages the advantages of Generative Adversarial Networks (GANs). Specifically, we propose a model that combines elements of the Pix2Pix model and the Wasserstein GAN (WGAN) with Gradient Penalty (WGAN-GP). This hybrid framework seeks to capitalize on the denoising capabilities of conditional GANs, as demonstrated in the Pix2Pix model, while mitigating the need for an exhaustive search for optimal hyperparameters that could potentially ruin the stability of the learning process. In the proposed method, the GAN's generator is employed to produce denoised images, harnessing the power of a conditional GAN for noise reduction. Simultaneously, the implementation of the Lipschitz continuity constraint during updates, as featured in WGAN-GP, aids in reducing susceptibility to mode collapse. This innovative design allows the proposed model to benefit from the strong points of both Pix2Pix and WGAN-GP, generating superior denoising results while ensuring training stability. Drawing on previous work on image-to-image translation and GAN stabilization techniques, the proposed research highlights the potential of GANs as a general-purpose solution for denoising. The paper details the development and testing of this model, showcasing its effectiveness through numerical experiments. The dataset was created by adding synthetic noise to clean images. Numerical results based on real-world dataset validation underscore the efficacy of this approach in image-denoising tasks, exhibiting significant enhancements over traditional techniques. Notably, the proposed model demonstrates strong generalization capabilities, performing effectively even when trained with synthetic noise.
\end{abstract}


\section{Introduction}\label{introduction}

\IEEEPARstart{T}{he digital} world is increasingly becoming inundated with vast amounts of visual data, much of which suffers from various forms of degradation, including noise \cite{fan2019brief,goyal2020image,asheghi2022comprehensive,oskouei2021cgffcm,kassem2023explainable}. Noise in images can severely compromise their quality and utility, posing significant challenges for critical tasks such as image recognition, object detection, and document digitization. Image denoising, therefore, represents a critical pre-processing step in many image-processing pipelines. Traditional image-denoising techniques often struggle with preserving image details, leading to a loss in high-frequency components and image blurring. Recently, Generative Adversarial Networks (GANs) have shown promise in various image processing tasks \cite{creswell2018generative, perez2023data}, including image denoising, due to their ability to learn high-level features from data. More advanced variants of GANs have rapidly come out. In \cite{wang2022adaptive} the authors proposed an adaptive self-adjusting learning for the generator network, to iteratively refine the denoised images. In the work of \cite{vo2021hi}, the authors employed a sophisticated architecture, comprehensive of three generator networks, to preserve both high-frequency and low-frequency components in the images. In this context, the Pix2Pix model \cite{isola2017image}, a type of conditional GAN, has demonstrated notable capabilities in tasks such as image-to-image translation, including noise reduction. The Pix2Pix model is particularly effective due to its ability to conditionally generate outputs based on input data, making it suitable for structured noise removal tasks. However, training GANs can be an unstable process, often leading to 'mode collapse,' where the generator produces a limited variety of samples. The Wasserstein GAN with Gradient Penalty (WGAN-GP) \cite{adler2018banach} presents a solution to this problem by introducing a Lipschitz continuity constraint during updates, thus improving the training stability. WGAN-GP enhances GAN training by providing more stable and reliable convergence, making it a valuable improvement for complex image processing tasks.\\
Binary images are common in document digitization, where old documents need to be scanned and pre-processed. When denoising a binary image, we are dealing with sharp transitions between the pixel values and less data since the data is quantized, which means the key challenge is to remove noise without eroding the important features of the image, and without having relevant frequency-domain information available to us \cite{li2021assessing,saxena2014noises}. This challenge is particularly pronounced in binary images, as the quantized nature of binary data demands precise noise removal to maintain the integrity of document features.
In the proposed study, an innovative hybrid architecture that blends the Pix2Pix model's denoising capabilities with the WGAN-GP's enhanced stability properties is proposed, aiming to produce denoised images while preserving the main key features present in binary images. This hybrid approach seeks to leverage the strengths of both models to achieve superior denoising performance, addressing the unique challenges posed by binary images and enhancing the overall quality of digitized documents.

\subsection{Motivations}
Binary image denoising is a critical task in various applications, including document processing, medical imaging, and pattern recognition. Traditional denoising methods, such as morphological operations, connected component analysis, rule-based filtering, simulated annealing, and median filtering, often fall short when effectively handling complex noise patterns and preserving fine structural details in binary images. Furthermore, advanced methods like the Ising model and simulated annealing, though effective, are computationally intensive and less adaptable to diverse noise patterns.
	In recent years, GANs (Generative Adversarial Networks) have shown great promise in image-denoising tasks due to their ability to learn complex distributions and generate high-quality images. GANs consist of two neural networks, the Generator, and the Discriminator, that compete against each other. The Generator creates denoised images from noisy inputs, while the Discriminator attempts to distinguish these generated images from real clean images. This adversarial setup allows the Generator to improve its denoising capabilities over time, driven by the feedback from the Discriminator.
	One significant advantage of GANs over traditional autoencoders is the presence of the Discriminator, which acts as a learned loss function, providing a more sophisticated evaluation of the denoised outputs than simple pixel-wise losses like Mean Squared Error (MSE) or Binary Cross-Entropy. This leads to the generation of denoised images that are more visually and structurally accurate.
	However, training GANs can be challenging due to mode collapse and instability. To address these challenges, we propose a novel hybrid model that combines the strengths of Pix2Pix and WGAN-GP (Wasserstein GAN with Gradient Penalty). Pix2Pix, a conditional GAN architecture, is well-suited for image-to-image translation tasks and has shown effectiveness in denoising applications. By incorporating residual blocks in the bottleneck, as demonstrated in \cite{jadhav2022Pix2Pix}, we can mitigate issues related to gradient vanishing and explosion.
	WGAN-GP introduces a Lipschitz continuity constraint that enhances training stability and reduces susceptibility to mode collapse. Our hybrid model seamlessly merges the structural preservation capabilities of Pix2Pix with the stability enhancements of WGAN-GP, functioning as a single, cohesive framework. This combination is particularly effective for binary images, where maintaining precise structural details such as text and graphical elements is crucial.
	In summary, our approach addresses the limitations of existing methods by offering a computationally efficient and highly effective denoising solution specifically tailored for binary images. The hybrid Pix2Pix WGAN-GP model not only preserves fine structural details but also ensures stable and robust training, making it a practical and scalable solution for various binary image denoising applications.

\subsection{Paper Contributions \& Structure}
\begin{figure}[]
	\centering
	\includegraphics[scale=0.25]{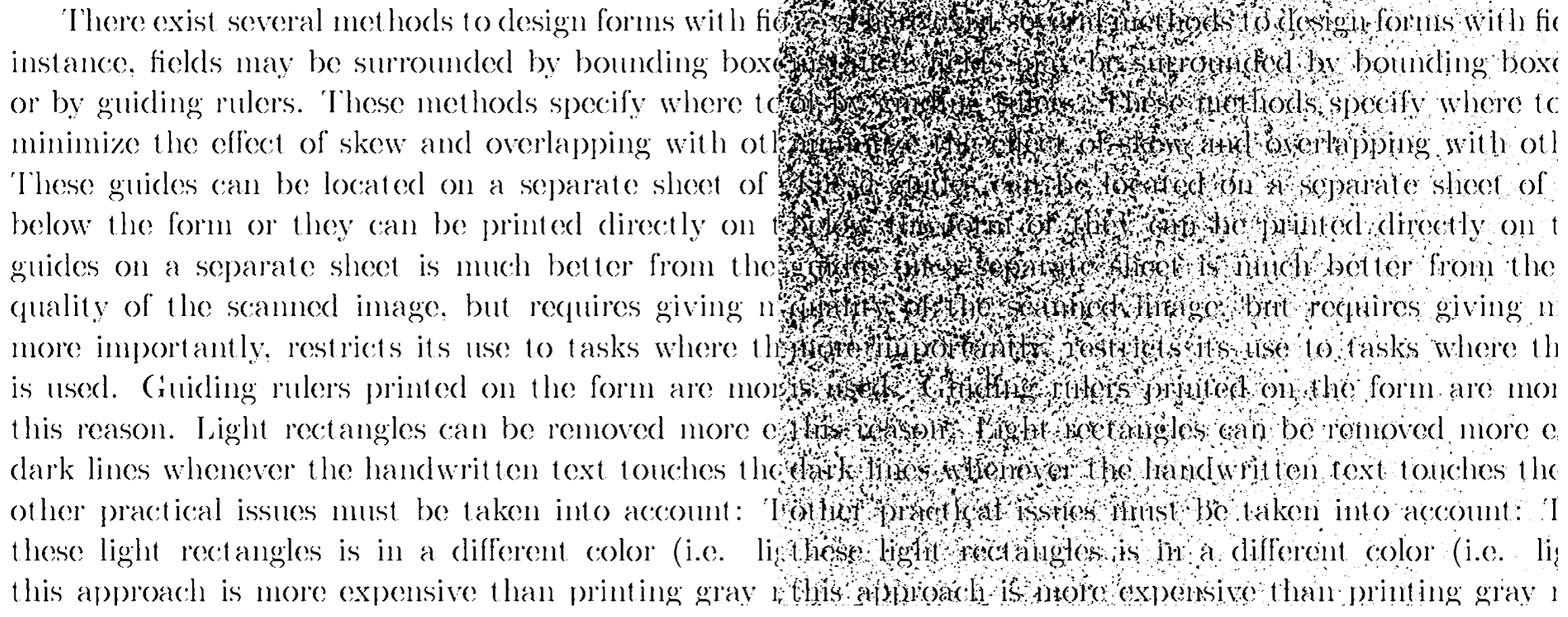}
	\caption{An Example of a clean-noisy couple in the Dataset.}
	\label{fig:dataset}
\end{figure}

The main contributions of this work can be summarized in the following points:
\begin{itemize}
	
	\item A comprehensive binary image dataset tailored for denoising tasks has been developed. This dataset comprises clean-noisy pairs (see Fig.\ref{fig:dataset}) generated through binarization techniques and the addition of different synthetic noise patterns, in terms of noise type, intensity, localization, and density.
	\item Unlike \cite{jadhav2022Pix2Pix} and \cite{sun2022pix2pix}, which are susceptible to mode collapse, a hybrid Pix2Pix WGAN-GP framework is proposed. This innovation aims to enhance the training stability of GANs, addressing the notorious challenges faced during standard GAN training.
	\item The induction of mode collapse in the training process is achieved through extensive hyperparameter search. Through this rigorous exploration, we observed mode collapse in the classical model, underscoring the resilience of the hybrid approach.
\end{itemize}

\paragraph*{Structure} The remainder of the paper
is organized as follows: Section \ref{sec:Related Works} presents the main works in the literature about binary image denoising and related works about conditions for model collapse. Section \ref{sec:Proposed Hybrid GAN Model} introduces the proposed hybrid GAN model and outlines the architecture and loss functions formulation. Section \ref{sec:Results}
illustrates the effectiveness of the proposed framework through numerical experiments and highlights a case of mode collapse. Finally, Section \ref{sec:Conclusions} concludes
the work.

\subsection{Preliminaries \label{sec:Preliminaries}}
In this paper, $\mathbb{E}$ denotes the expected value of an element, $\mathbb{E}_{x \sim y}$ refers to the conditional expected value of $x$ given $y$, $\text{Var}$ describes the variance, $\nabla$ denotes the gradient operator, $\theta_{\text{gen}}$ and $\theta_{\text{dis}}$ are the neural network parameters, $G(x)$ describes the Generator model, and $D(x, x_{\text{noisy}})$ denotes the Discriminator model. $x_{\text{clean}}$ refers to clean input image, $x_{\text{denoised}}$ describes a denoised image, $x_{\text{noisy}}$ denotes a noisy input image, and $x_{\text{interp}}$ expresses the interpolation between clean and denoised samples.

\subsubsection{Conditions for Mode Collapse}

A prevalent issue in the training of GANs is mode collapse, a phenomenon where the generator starts producing a narrow range of outputs, regardless of the diversity of inputs it receives. Understanding the mathematical underpinnings of mode collapse can be essential for developing strategies to prevent it. 
Mode collapse occurs when the generator learns to exploit the weaknesses in the discriminator, finding specific paths in the parameter space that achieve high scores from the discriminator but do not correspond to meaningful or diverse solutions. This phenomenon can be observed mathematically through the behavior of the cost function \(J(D, G)\) used to train the discriminator and the generator.
Firstly, the generator's output distribution \(P_G\) and the real data distribution \(P_R\) are modeled. The goal is generally to minimize the distance between these distributions, which can be represented by various metrics such as Jensen-Shannon divergence or Wasserstein distance \cite{cai2022distances}. Mode collapse tends to occur when this distance metric is minimized in a manner that does not encourage diversity in \(P_G\). 
Ideally, the objective of the generator can be expressed as follows:
\begin{equation}
	\begin{split}
		\min_G \max_D J(D, G) = \mathbb{E}_{(x_{\text{clean}}, x_{\text{noisy}}) \sim P_{R}}[\log D(x_{\text{clean}}, x_{\text{noisy}})] \\ 
		+ \mathbb{E}_{(x_{\text{denoised}}, x_{\text{noisy}}) \sim P_{G}}[\log(1 - D(x_{\text{denoised}}, x_{\text{noisy}}))] 
	\end{split}
\end{equation}
where $(x_{\text{clean}}, x_{\text{noisy}})$ denotes a real couple in the dataset that belongs to $P_R$, and $(x_{\text{denoised}}, x_{\text{noisy}})$ is the fake couple, belonging to the distribution of the generator's output $P_{G}$. In this equation, the first term represents the expected value of the log-probability that the discriminator assigns to the real, clean-noisy pairs, and the second term represents the expected value of the log-probability that the discriminator assigns to the fake, denoised-noisy pairs generated by $G$.
During mode collapse, the generator starts optimizing to produce a limited set of outputs that highly deceive the discriminator. This situation can lead to a scenario where the gradient of the cost function with respect to the generator's parameters becomes almost null, indicating that the generator is no longer learning to produce diverse outputs. A good metric for this is analyzing the variance in the gradient of the cost function with respect to the generator parameters, denoted as \( \text{Var}\left(\nabla_\theta J(D, G)\right) \). A decreasing variance indicates that the generator is focusing on a smaller set of solutions, which can be a precursor to mode collapse. In mathematical terms, mode collapse is imminent when: $\text{Var}\left(\nabla_\theta J(D, G)\right) \rightarrow 0$. Here, \( \theta \) are the generator's parameters. The Nash equilibrium in this game is a point where both the generator and the discriminator cannot further decrease their respective loss functions. However, achieving this equilibrium is highly challenging, and mode collapse can be seen as a failure to reach a stable equilibrium where \(P_G\) approximates \(P_R\) diversely and accurately \cite{thanh2020catastrophic}. To mitigate mode collapse, it is essential to develop training strategies that encourage the generator to explore a larger portion of the parameter space, promoting more diverse outputs. In subsequent sections, various strategies will be implemented on the proposed hybrid model to prevent mode collapse and foster a more stable and beneficial training process. Moreover, in the specific case addressed by this work, the images are characterized by reduced information content due to their binary nature; this, in the context of denoising, may induce mode collapse during training, especially if the learner has to deal with local information only, such as the adoption of a tailing procedure. This happens because, during training, many noisy-clean pairs might be the same white, noise-free patch (because not all the patches of the processed image may be dirty), leading the generator to produce fully white output patches, limiting the diversity of its output.

\section{Related Works \label{sec:Related Works}}
\subsection{Literature Review}
Binary image denoising tasks are still a challenging open problem, with a wide range of techniques where state-of-the-art denoising approaches have several advantages and disadvantages \cite{fan2019brief,buades2005review,saba2014evaluation}. Techniques such as median filtering and morphological operations like erosion and dilation, which are particularly effective at preserving edges and structures, are often used \cite{jamil2008noise,hashemzadeh2019content} The main drawbacks of these approaches are that they lack a mechanism for identifying, distinguishing, and preserving key structural components present in the images, and they depend heavily on the choice of parameters (like filter size). This often necessitates tuning procedures that work well only if the images in the dataset are homogeneous, which may not always be the case and may remove the key features \cite{fan2019brief}. Other techniques include Connected Component Analysis, where each connected component in the image is isolated, analyzed, and then filtered based on some predefined rules like a compactness threshold for the components \cite{salimi2014automatic,preeti2020denoising}. This technique is powerful when the images present a very regular noise profile, allowing for the removal of noise components based on some heuristics, but it is poorly robust against outliers (like the points on top of "i" letters being detected as noise) due to the difficulty of labeling these components \cite{he2017connected}. Moreover, this can be computationally expensive if many components are present in the images.
\\
The recent developments in the field of Artificial Intelligence (AI) and Machine Learning (ML) have resulted in useful methodologies that are used for many applications such as wildfire \cite{jonnalagadda2024segnet}, cast defects \cite{chamberland2023autoencoder}, and efficient denoising while preserving structural features in the images \cite{izadi2023image}. Residual learning is a technique widely used in literature \cite{eltoukhy2022classification}, and can be used for denoising too, as have been done through DnCNN approach, where a deep learning model is used to predict the residual noise given a noisy input image \cite{zhang2017beyond}. The structure of the network ensures that, if the residual noise is properly estimated, it can be subtracted from the input noisy image to yield a denoised image. Another common approach is to use Autoencoders and some of their variants like U-NETs. They work by encoding the input data into a compressed representation, passing it through a bottleneck, and then decoding this representation back into the original format \cite{vincent2008extracting,majumdar2018blind}.\\
Denoising autoencoders are variants specifically designed to use the reconstruction process to remove noise from input data. They are trained to reconstruct the noise-free data from a noisy version of the data. In the literature, Variational Autoencoders (VAEs) have also gained popularity, providing a probabilistic manner for describing an observation in latent space. Therefore, instead of mapping an input to a fixed vector, VAEs map the input to a distribution. When used for image denoising, VAEs can be utilized to model the distribution of clean images, and given a noisy image, VAEs are capable of generating a clean image drawn from the same distribution \cite{kingma2013auto} \cite{im2017denoising}. Previous works on GANs showed the versatility of such approaches in various image-processing tasks. Drawing inspiration from \cite{chen2018image}, where the authors proposed a blind denoising method combining GANs with CNNs, the authors of \cite{tran2020gan} have used GANs for realistic image noise modeling, estimating noise intensity, and coupled this strategy with more advanced denoising methods, such as DnCNN, to robustly remove noise from images. Wang et al. further advanced this domain by integrating multifaceted loss functions to enhance detail retention in denoised images \cite{wang2020image}. Their work introduces a Deep Residual GAN (DRGAN) designed to not only refine the visual quality of denoised images but also extend its applicability to other image processing tasks such as defogging and medical image enhancement.

\subsection{Training Instability and Mode Collapse }
As previously mentioned, the instability in GANs' training process is still a challenging open problem. The dual nature of the training introduces coupled dynamics between the generator and discriminator. The challenge of coupled dynamics could potentially lead to one generator outperforming the discriminator and vice versa \cite{saxena2021generative}. As a result, GANs are subject to mode collapse, which typically happens when the generator finds a "wrong" shortcut to fool the discriminator. This shortcoming results in losing generalization capabilities and produces the same or similar output no matter what input is fed to it \cite{chen2021challenges}.
To address this critical challenge, Wasserstein GANs (WGANs) are adopted, employing a different way of computing the losses and updating the models \cite{gulrajani2017improved, ghasemieh2023enhanced}. In particular, the authors tried to render the updates of the discriminator as smoothly as possible, by first introducing weight clipping and later evolving to a gradient penalty term in the loss of the discriminator (WGAN-GP), similar to what was done in Reinforcement Learning with \cite{schulman2017proximal,ali2024deep,ali2023action}. The authors of \cite{chu2020smoothness} tried to develop a solid mathematical understanding of the stability criteria behind GANs' training processes and provided some regularity conditions on several frameworks, highlighting superior stability properties (not guaranteed) in the WGAN-GP case. The authors of \cite{thanh2019improving} proposed a zero-centered gradient penalty to improve the generalization capabilities of WGANs. In \cite{thanh2020catastrophic} the authors investigated the relation between catastrophic forgetting, non-convergence, and mode collapse. They concluded that methods like imbalanced loss, zero-centered gradient penalties, optimizers with momentum, and continual learning are effective at preventing catastrophic forgetting in GANs. Furthermore, the work in \cite{thanh2020catastrophic} emphasized that the gradient penalty adoption may improve the generalization capabilities of the network.

\section{Proposed Hybrid GAN Model \label{sec:Proposed Hybrid GAN Model}}
\label{sec:proposedmodel}

\subsection{Motivation}
\begin{figure}[!htbh]
	\begin{centering}
		\includegraphics[width=0.5\textwidth]{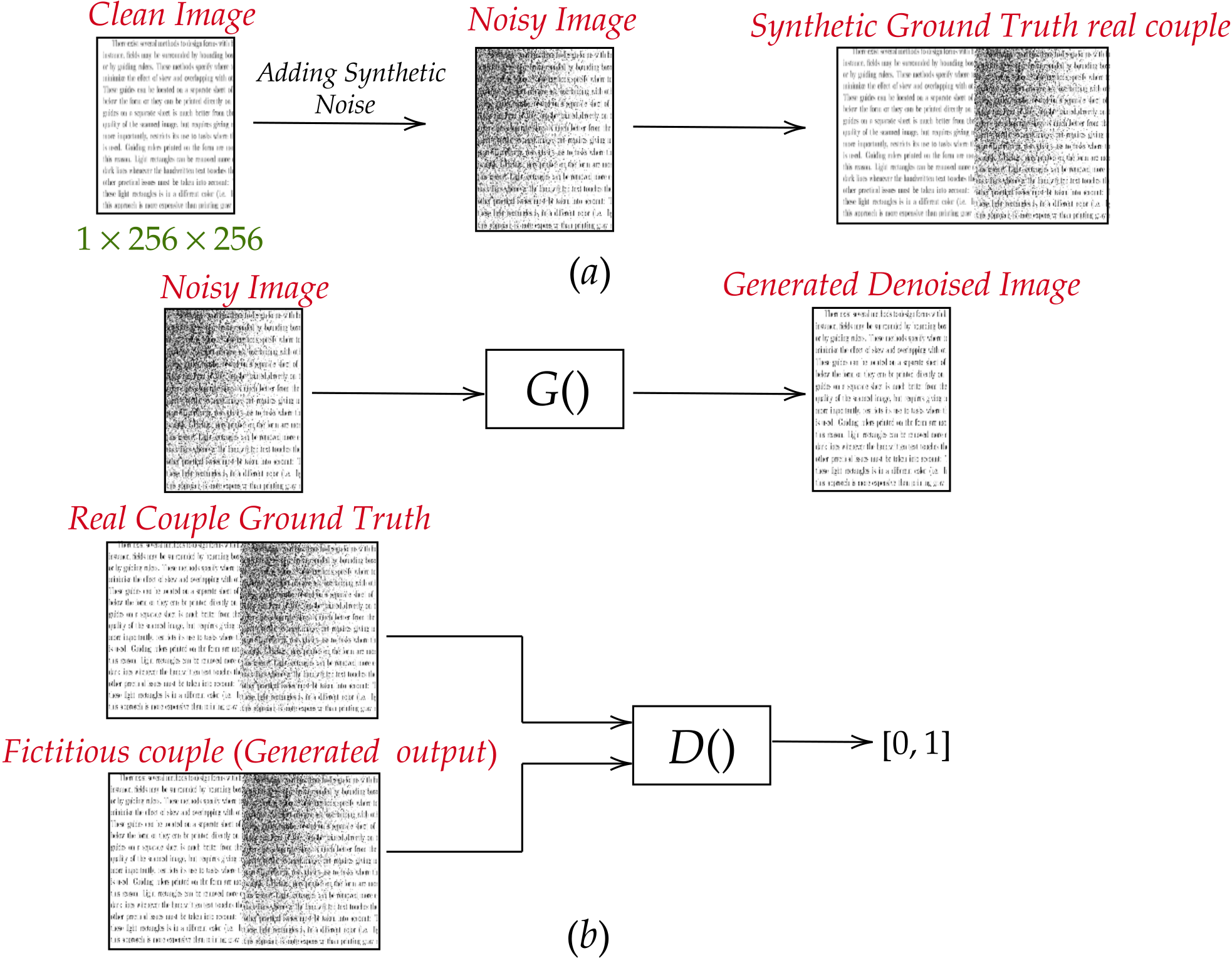}
		\par\end{centering}
	\caption{Introductory scheme of the framework. Part (a) shows the clean, noisy image after adding synthetic noise, and the synthetic ground truth real couple Part (b) illustrates the inputs and outputs of both generator and discriminator models. Discriminator's goal is to declare if the clean image in the couple is a real or generated image.}
	\label{fig:intro}
\end{figure}

In the novel proposed GAN framework, two primary components are employed: a generator and a discriminator. The generator aims to translate a noisy image into a clean image. A model working at the patch level to handle variable-sized images is adopted. The images are split into patches that reassemble the original image after being denoised. The discriminator is only used to train the generator. It aims to distinguish between pairs of clean-noisy and denoised-noisy images. The architecture of the proposed hybrid model, is the same as the architecture of classical Pix2Pix, taken as a baseline; what changes is how we compute the losses for training these models. In Fig. \ref{fig:intro}, a scheme illustrating the problem formulation is shown to introduce the reader to the Pix2Pix, used as an evaluation baseline for our hybrid denoising framework.

\subsection{Network Architectures}
In the classical pix2pix and in the hybrid pix2pix WGAN-GP models, the generator and discriminator architectures are identical. The difference lies in the structure of their loss functions.
\subsubsection{Generator Architecture}
\begin{figure*}[!htbh]
	\begin{centering}
		\includegraphics[width=1\textwidth]{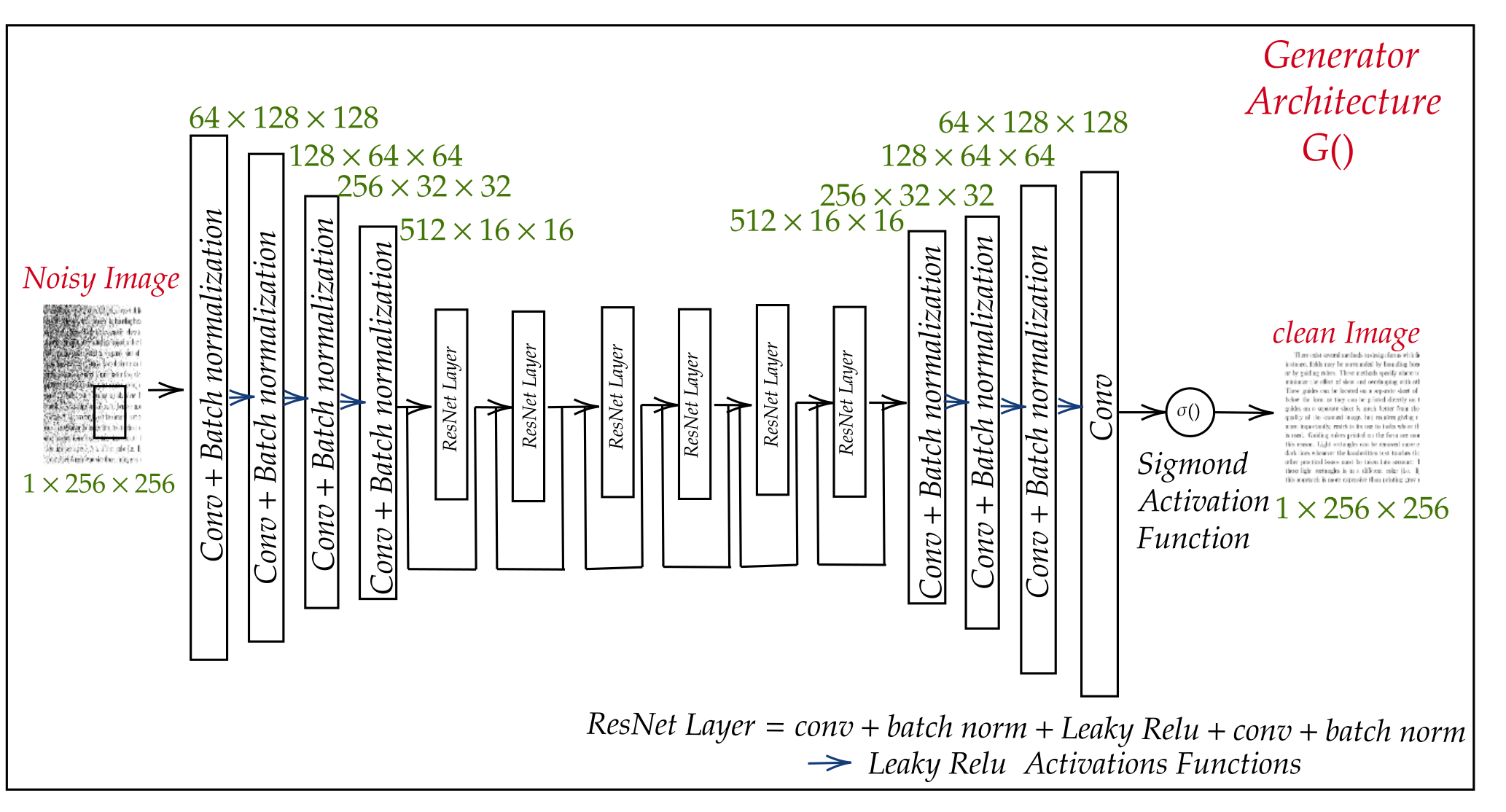}
		\par\end{centering}
	\caption{Generator Autoencoder Architecture with ResNet Blocks.}
	\label{fig:gen_network}
\end{figure*}

The generator network, indicated with the $G$ symbol, takes as input a noisy patch of size $256\times 256$ and outputs a clean patch of the same size. It is a classical autoencoder model that consists of an encoder block, a pass, and a decoder. Fig. \ref{fig:gen_network} illustrates the detailed architecture of this network.
Following the approach of \cite{jadhav2022Pix2Pix}, a pass consisting of a series of six ResNet blocks using 2D-Convolution is adopted, with Batch Normalization, Leaky ReLU activation functions, 2D-Convolution, and a Batch Normalization layer. This has the advantage of reducing the effect of the vanishing or exploding gradients and in our case, it worked better than a U-NET architecture \cite{zhou2018unet++}. The encoder and decoder blocks are composed of several blocks using 2D Convolution (transposed in the decoder), batch normalization, and Leaky ReLU. Since we are dealing with binary images, hyperbolic tangent with rescaling or Sigmoid activation functions were tested at the output layer. The convolutional layers are used to extract the core and dominant features from the input images, mapping them to a latent space while discarding noise. The deconvolution that operates in the decoder part aims to reconstruct the real images without noise, based on the features captured in the encoder.

\subsubsection{Discriminator Architecture}
\begin{figure*}[!htbh]
	\begin{centering}
		\includegraphics[width=1\textwidth]{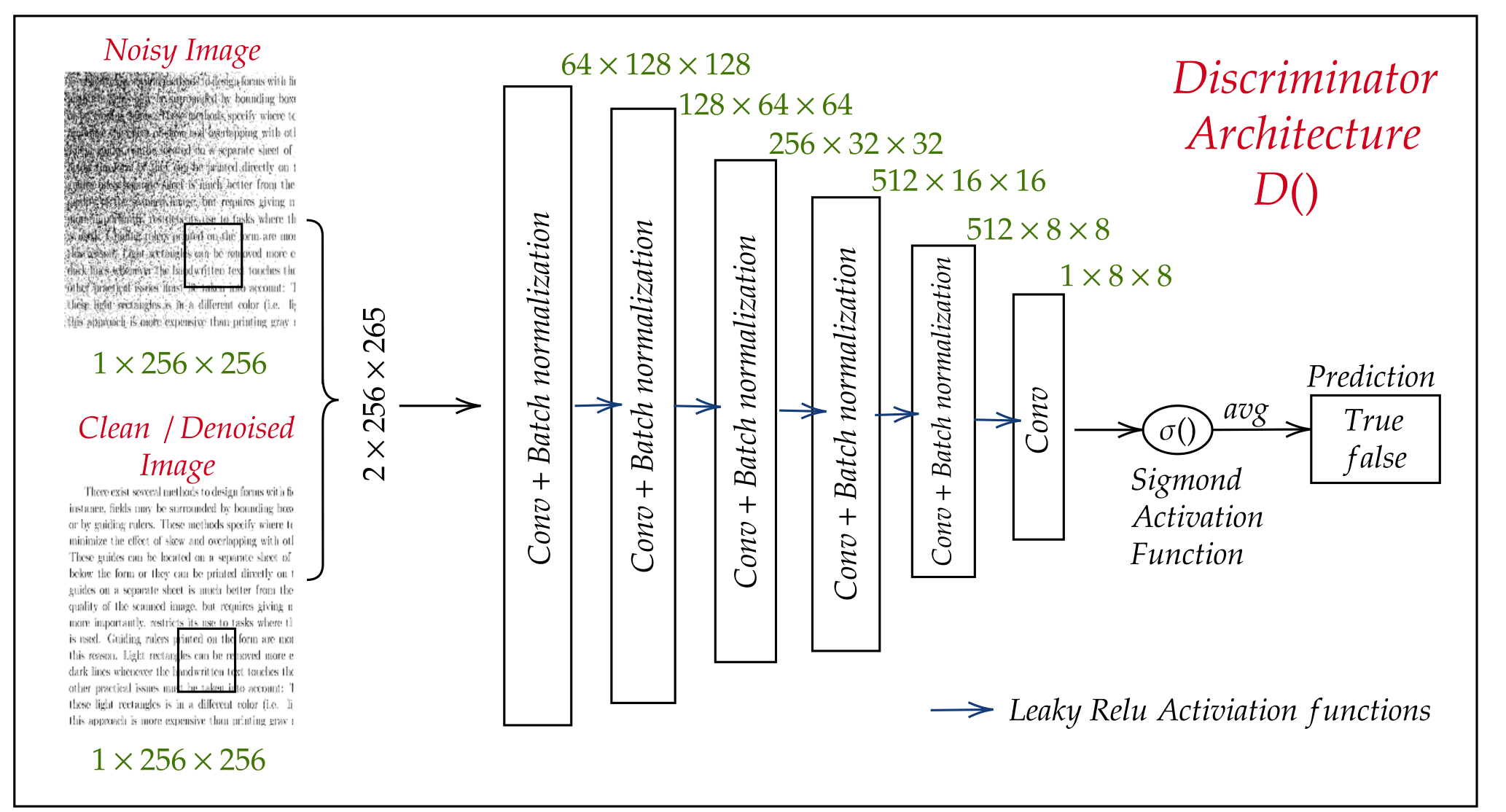}
		\par\end{centering}
	\caption{Discriminator PatchGAN Architecture.}
	\label{fig:discriminator_model}
\end{figure*}

The discriminator model is a conditional PatchGAN, indicated as D. It takes as input pairs of images, concatenates them along the channel dimension, and downsamples the image by outputting a smaller patch after some convolution, batch normalization, and ReLU blocks. The output $8\times 8$ patch is then averaged along width and height to get a scalar value for the model prediction on the pair, which is used for comparison with a ground truth binary label that indicates real and fictitious pairs. Fig. \ref{fig:discriminator_model} provides a detailed visualization of the discriminator's architecture. The learning phase aims to let the generator improve in the denoising process, fooling the discriminator into predicting the generated denoised-noisy fictitious pairs are real pairs.

\subsection{Loss Functions}
In both the proposed schemes, a well-known loss function is employed, the Binary Cross Entropy (BCE), which is defined as:
\begin{equation}
	\text{BCE}(y, \hat{y}) = -\frac{1}{N} \sum_{i=1}^{N} \left[ y_i \log(\hat{y}_i) + (1 - y_i) \log(1 - \hat{y}_i) \right]
\end{equation}

To improve numerical stability, the sigmoid activation function in the output layer of the network (\(\sigma(x) = \frac{1}{1+e^{-x}}\)) is integrated into the BCE loss function. Consider \( \hat{y} = \sigma(x) \) such that the BCE loss function with integrated sigmoid is then given by:
\begin{equation}
	\begin{aligned}
		\text{BCE}(y, x) = & -\frac{1}{N} \sum_{i=1}^{N} \bigg[ y_i \log\left(\sigma(x_i)\right) \\
		& \hspace{60pt} + (1 - y_i) \log\left(1 - \sigma(x_i)\right) \bigg]
	\end{aligned}
\end{equation}
where $y$ are the true labels, $x$ are the predicted labels before applying the sigmoid function and $N$ represents the number of samples.

\subsubsection{Classical Pix2Pix Model}
The following scheme summarizes the classical Pix2Pix loss formulation used in the implementation as a baseline, for an input patch of size $64\times 64$. For bigger patches, more encoding-decoding layers are added to the network to match the same kind of dimensional transition that happens in the convolutional blocks.

\paragraph{Generator}
In the classical Pix2Pix model, the loss of the generator consists of two terms: an adversarial loss and a regularization term.
The adversarial component of the loss is derived by comparing the output of the discriminator for denoised-noisy pairs with a label falsely indicating they are real couples, using a Binary Cross Entropy (BCE) with integrated Sigmoid activation function to improve numerical stability. The regularization term consists of a constant multiplied by the L1 loss between the denoised patches and the real clean ones. This regularizer improves the model's capabilities to produce images more similar to the clean ones.
The adversarial loss is computed as follows:
\begin{equation}
	J_{1, \text{adv}}(G, D) = BCE( D( x_{\text{denoised}}, x_{\text{noisy}} ), y_{\text{true}} )
\end{equation}
where $x_{\text{denoised}} = G(x_{\text{noisy}})$ and $y_{true}$ is a vector of ones of the size of the batch. The generator loss is then given by:
\begin{equation}
	J_{1,\text{gen}}(G, D) = J_{1,\text{adv}} + \lambda * L1( x_{\text{denoised}}, x_{\text{clean}} )
\end{equation}
enforcing the denoised patches to be similar to clean ones.

\paragraph{Discriminator}
The classical discriminator loss is computed by averaging the Binary Cross Entropy losses on the discriminator predictions for real and fictitious pairs.
The equations defining the Pix2Pix discriminator loss are as follows:
\begin{equation}
	J_{1,\text{real}}(D) = BCE ( D ( x_{\text{clean}}, x_{\text{noisy}} ) , y_{\text{true}} )
\end{equation}
\begin{equation}
	J_{1,\text{fake}}(G, D) = BCE ( D ( x_{\text{denoised}}, x_{\text{noisy}} ) , y_{\text{false}} )
\end{equation}
\begin{equation}
	J_{1,\text{dis}}(G, D) = \frac{J_{1,\text{real}} + J_{1,\text{fake}}}{2}
\end{equation}

\subsubsection{WGAN-GP Pix2Pix Hybrid Model}

\begin{figure*}[!htbh]
	\begin{centering}
		\includegraphics[width=1\textwidth]{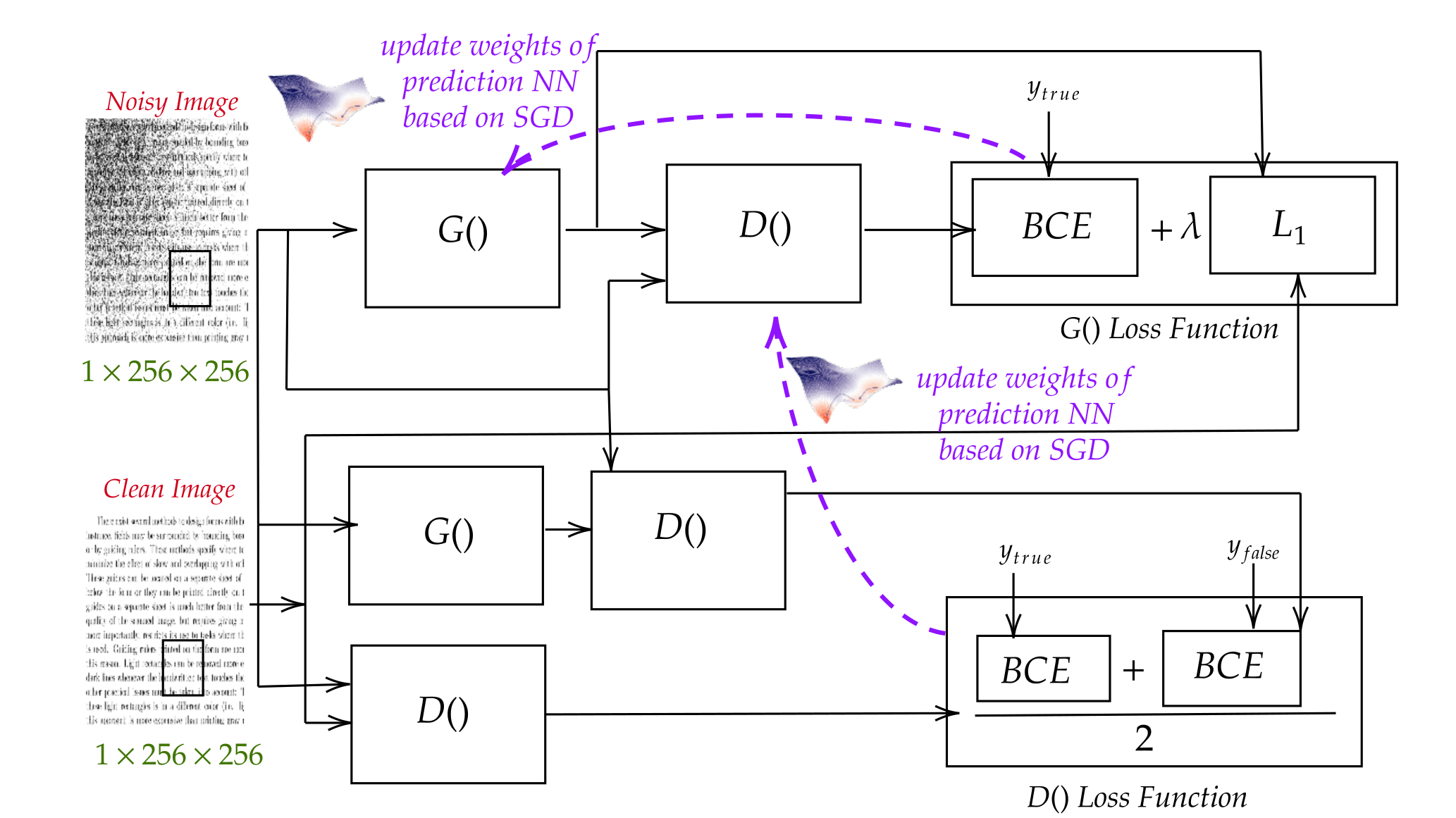}
		\par\end{centering}
	\caption{Hybrid Denoising GAN Framework Scheme}
	\label{fig:hybrid_scheme}
\end{figure*}

The original Wasserstein GAN (WGAN) uses the Wasserstein distance as an objective function, also known as the Earth Mover's distance. This distance metric measures the amount of 'work' needed to transform one probability distribution into another. In the context of GANs, this translates to determining how much effort is required to transform the generator's output distribution to match the real data distribution. In the original WGAN, weight clipping is proposed. However a more sophisticated version of the WGAN was recently introduced, making use of Gradient Penalty. We adopted this technique in the proposed approach. Fig.\ref{fig:hybrid_scheme} illustrates the Hybrid Denoising GAN Framework Scheme employed in this study.

\paragraph{Generator}
The discriminator in a WGAN is often referred to as the critic because, unlike traditional GANs, it does not try to classify inputs as real or fake. Instead, it is trained to output a scalar value for each input that roughly corresponds to the input's "realness". The more positive the value, the more "real" the input is considered, and the more negative the value, the more "fake" it is considered.
The generator's goal is to make its generated images appear as real as possible to the discriminator; it aims for the discriminator to output a positive value when evaluating a generated image, which is equivalent to minimizing the loss with the opposite sign.
To compose the adversarial term in the generator's loss using the WGAN, the negative average across the width and height dimensions of the discriminator's output is taken for a denoised-noisy pair.
\begin{equation}
	J_{2,\text{adv}}(G, D) = - D( x_{\text{denoised}}, x_{\text{noisy}} )    
\end{equation}
When working with a batch size, this term should be averaged across the entire batch  during training. The regularization term is still present and computed in the same way as before, with the L1 Loss.
\begin{equation}
	J_{2,\text{gen}}(G, D) =  J_{2,\text{adv}} + \lambda * L1( x_{\text{denoised}}, x_{\text{clean}} )
\end{equation}

\paragraph{Discriminator}
The discriminator loss in this approach is composed of two terms: the first component uses a similar strategy to the average used before, but involves a difference between real and fake predictions. The second term is used to enforce the gradient penalty.
\begin{equation}
	J_{2,\text{diff}}(G, D) =  D( x_{\text{clean}}, x_{\text{noisy}} ) - D( x_{\text{denoised}}, x_{\text{noisy}} )
\end{equation}
The latter component should also be averaged over the batch size if training works with batches.
The gradient penalty is a regularization term in the loss function that encourages the gradient norm of the discriminator's output with respect to its input, which in this case are interpolated samples between real and fake ones, to be close to one. It's computed as the average of the square of the difference between the gradient norm and one. The gradient penalty term is computed by first defining $x_{\text{interp}}$ as an interpolation between clean and denoised sample, as:
\begin{equation}
	x_{\text{interp}} = \alpha * x_{\text{clean}} + (1 - \alpha) * x_{\text{denoised}}
\end{equation}
where $\alpha \in (0,1)$ is a random number. Then, the gradient penalty is defined as :
\begin{equation}
	J_{2,gp}(G, D) = E[(||\nabla_{x_{interp}} D(x_{interp}, x_{noisy})||_2 - 1)^2]
\end{equation}
These two terms contribute to the definition of the discriminator's loss, which is defined as:
\begin{equation}
	J_{2,dis}(G, D) = J_{2,diff} + w_{p} * J_{2,gp}
\end{equation}
where $w_p$ is a weight for the gradient penalty mentioned in Table \ref{tab1}. The pseudocode of the hybrid model training loop is presented in Algorithm \ref{alg:Pix2PixDenoising}. In the algorithm, the $PreProcess()$, $GetBatch()$, $GetInterpolation()$, $GetGradPenalty()$ and $Update()$ functions are used to simplify the pseudocode. The first function performs some preprocessing operations typically used with images, such as padding to a uniform size, random cropping, and random rotations. The other three are used to specify batches from the dataset, perform interpolation, compute gradient penalty, and update network weights.

\begin{algorithm}
	\caption{WGAN-GP Hybrid Training for Image Denoising} \label{alg:Pix2PixDenoising}
	
		\textbf{Require} Noisy dataset $D_{\text{noisy}}$\\
\textbf{Require} Clean dataset $D_{\text{clean}}$\\
\textbf{Require} Generator $G$\\
\textbf{Require} Discriminator $D$\\
\textbf{Require} L1 Loss coefficient $\lambda_{L1}$\\
\textbf{Require} Gradient Penalty coefficient $w_p$

	\textbf{for $n_{\text{epochs}} = 1, \dots , N_{\text{epochs}}$}
	\begin{enumerate}
		\item[{\footnotesize{}1:}]  $x_{\text{clean}}, x_{\text{noisy}} = PreProcess(x_{\text{clean}}, x_{\text{noisy}})$
		\item[{\footnotesize{}2:}]  $x_{\text{clean}}, x_{\text{noisy}} = GetBatch(D_{\text{clean}}, D_{\text{noisy}})$
		\item[{\footnotesize{}3:}]  $x_{\text{denoised}} = G(x_{\text{noisy}})$
		\item[{\footnotesize{}4:}]  $x_{\text{interp}} = GetInterpolation(x_{\text{clean}}, x_{\text{denoised}})$
    	\item[{\footnotesize{}5:}]	$D_{\text{real}} = D(x_{\text{clean}}, x_{\text{noisy}})$
    	\item[{\footnotesize{}6:}]	$D_{\text{fake}} = D(x_\text{denoised},x_\text{noisy})$
    	\item[{\footnotesize{}7:}]	$J_{\text{real}} = BCE(D_{\text{real}}, \text{ones})$
    	\item[{\footnotesize{}8:}]	$J_{\text{fake}} = BCE(D_{\text{fake}}, \text{zeros})$
    	\item[{\footnotesize{}9:}]  $J_{\text{diff}} = D_{\text{real}} - D_{\text{fake}}$
		\item[{\footnotesize{}10:}] $J_{gp} = GetGradPenalty(x_{\text{interp}}, x_{\text{noisy}})$
		\item[{\footnotesize{}11:}] $J_{dis} = J_{\text{diff}} + w_p \times J_{gp}$
		\item[{\footnotesize{}12:}] Update $D$ minimizing $J_{\text{dis}}$
	\item[{\footnotesize{}13:}] $J_{\text{adv}} = -D_{\text{fake}}$
	\item[{\footnotesize{}14:}] $J_{\text{gen}} = J_{\text{adv}} + \lambda_{L1} \times L1Loss(x_{\text{denoised}}, x_{\text{noisy}})$
	\item[{\footnotesize{}15:}] Update $G$ minimizing $J_{\text{gen}}$
	\item[{\footnotesize{}16:}]\textbf{if} {$n_{\text{epochs}} \mod f_{\text{save}} = 0$}
	\item[{\footnotesize{}17:}] $Update(\theta_{\text{gen}}, \theta_{\text{dis}})$
		\item[{\footnotesize{}18:}]\textbf{end if}
	\end{enumerate}
	\textbf{end for}
\end{algorithm}

\subsection{Assumptions and Limitations}
Our proposed work considers only the case of binary images. Noise in these kinds of images shares characteristics with the main signal content, making it challenging to distinguish between the two. For example, noisy pixels share the same intensity or spatial distribution as the foreground or background objects in the image. The assumptions of the proposed work can be summarized as follows:
\begin{enumerate}
	\item Only binary images have been considered.
	\item Pix2Pix architecture relies on paired training data consisting of clean images and their corresponding noisy versions. The assumption here is that for every clean image, there exists a corresponding noisy version.
	\item The noisy images were constructed by adding synthetic noise to the clean images.
\end{enumerate}
The limitations of the proposed work can be summarized as follows:
\begin{enumerate}
	\item The need for a large amount of training data, possibly matching the real use case images on which the model should operate during inference.
	\item Imbalanced domains in practical scenarios.
	\item In real use case scenarios, the synthetic noise can be due to several different factors (analog interference during the scanning process, electronic noise in the scanner's imaging sensors, bleedthrough, book bindings not properly managed), and noise data must be generated according to the noise type and intensity.
	\item As mentioned, GAN mode collapse is an open issue, especially for long training processes. The adoption of Wasserstein Loss and the gradient penalty helps to achieve a more balanced training process, making it more difficult for the generator or discriminator to outperform the other.
\end{enumerate}

\section{Experiments \& Results \label{sec:Results}}

\subsection{Data Collection and Preprocessing}
The dataset used in this study consists of binary image pairs, specifically designed to include both clean and artificially generated noisy versions. The clean version of the images was sourced from Kaggle (\url{https://www.kaggle.com/c/denoising-dirty-documents/data?select=test.zip}). Since the clean images were in grayscale format, they were binarized. The binarization process was performed using a binary threshold combined with Otsu's algorithm. This dynamic thresholding automatically determines the optimal threshold value by analyzing the histogram of the image to find the threshold that minimizes the weighted within-class variance, effectively separating the pixel values into two groups: foreground and background. After binarization, each clean image underwent a controlled noise addition process to simulate various types of image degradation. This process was implemented using the Augraphy library, which is specifically designed for generating realistic noisy effects in digital images (\url{https://github.com/sparkfish/augraphy}). The noise model used was 'BadPhotoCopy', which allows for extensive customization of noise characteristics. The noise model was configured to randomly vary several parameters to ensure a diverse set of noisy images. These parameters included the type of noise, which could range from subtle distortions to severe degradation patterns; the intensity and sparsity of noise; and additional effects like blurring and edge distortions. Each image was processed with a unique set of parameters, chosen randomly within specified ranges. The noise was applied to the image in a non-uniform manner, affecting different parts of the image to varying degrees, which further enhanced the realism of the simulated noise. For example, noise iteration ranged between 2 and 5, noise size between 2 and 6 pixels, and noise values adjusted for brightness and contrast between 20 and 200. The script used to add noise to the clean images can be found in our GitHub repository. To enhance the diversity and volume of our dataset, each clean image was utilized to generate three distinct noisy versions. This approach allowed us to expand the dataset effectively by tripling the number of images available for training, while maintaining a consistent baseline of clean images for comparative purposes. \\
	The full training pipeline also included random cropping of the images to $1024\times 1024$, with padding if necessary. This was done to tackle variable-sized images, to improve the dataset's diversity, and to ensure a fixed batch size (variable-sized image decomposition may lead to a different number of patches per image). The images were then decomposed into smaller patches that were randomly rotated to improve generalization capabilities in the learning of core features. Then the patches were denoised one by one and reassembled to recompose the original image, cropping out any padding.

\subsection{Parameters and Hardware}
The model was trained for 100 epochs, over a duration of 4 hours and 52 minutes. The training was conducted on a Windows 11 OS, with an Intel(R) Core(TM) i7-10750H CPU @ 2.60GHz, 16.0 GB of RAM, and an NVIDIA GeForce GTX 1660 Ti, using 500 GB of SSD storage.
To provide a more in-depth analysis of the computational complexity of the GAN model, the Multiply-Accumulate Operations (MACs) were calculated. MACs represent the total number of operations required to process an input through the model, where each operation involves multiplying two numbers and then adding the result to an accumulator.
For this GAN model, the MACs for the generator network are 15.406 GMACs, while the discriminator network has a MAC value of 1.921 GMACs.
In Table \ref{tab1}, the parameters used for the optimal training process are summarized.

\begin{table}[hbt!]
	\begin{center}
		\caption{Agent Parameters}
		\begin{tabular}{| c | c | c |} 
			\hline
			\label{tab1}
			Hyperparameters & Classical Pix2Pix & WGAN-GP Pix2Pix \\ [0.5ex] 
			\hline \hline
			Learning Rate Generator & 0.0002 & 0.0002 \\ 
			\hline
			Learning Rate Discriminator & 0.0002 & 0.0002 \\ 
			\hline
			Batch Size (images) & 4 & 4 \\
			\hline
			Lambda & 30000 & 30000 \\
			\hline
			Gradient Penalty Weight & - & 10  \\
			\hline
			Weight Clamping Value & - & 0.01 \\
			\hline
		\end{tabular}
	\end{center}
	\vspace{-4mm}
\end{table}

\subsection{Results}
\begin{figure}[]
	\centering
	\includegraphics[scale=0.36]{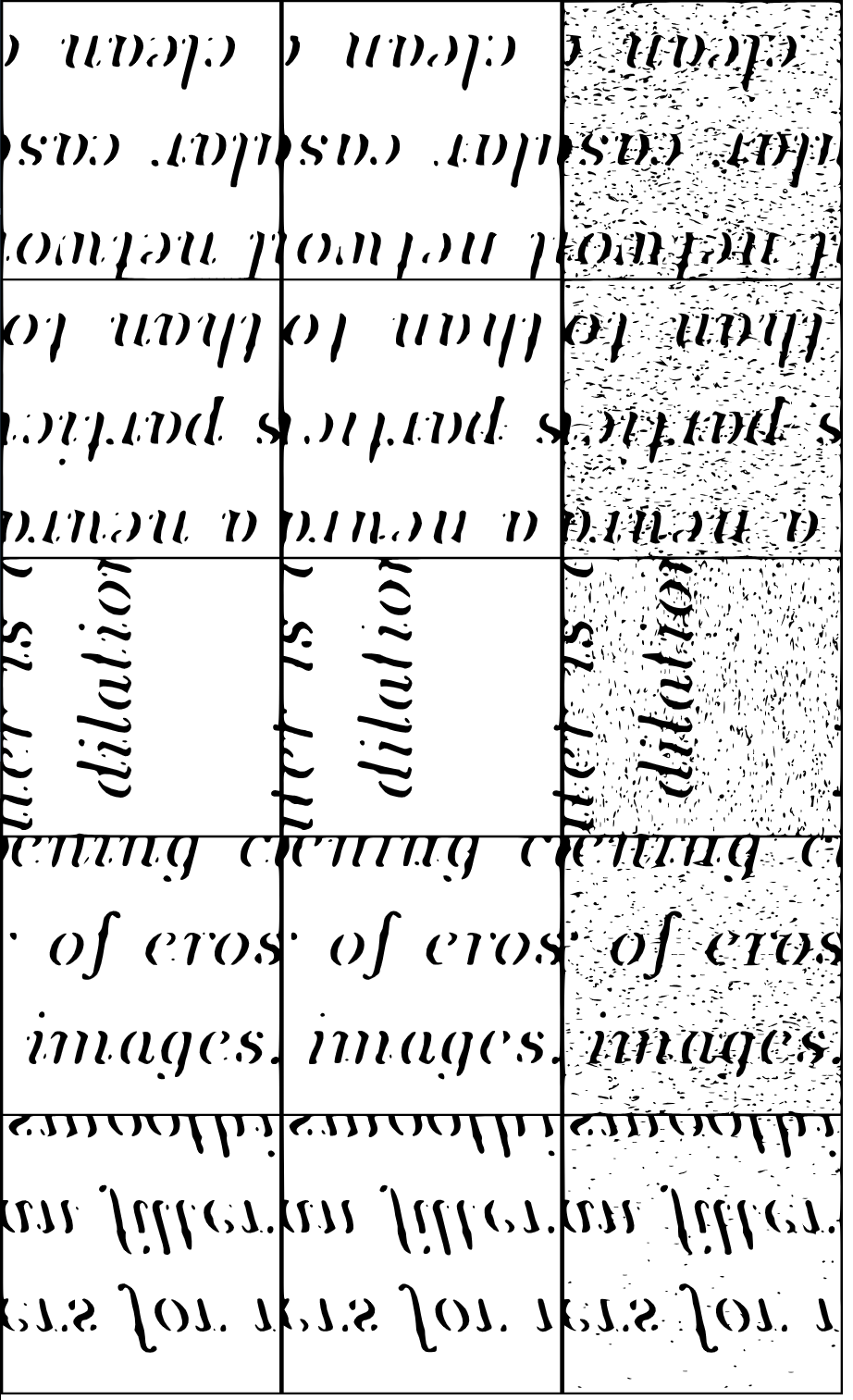}
	\caption{An Example of clean-denoised-noisy triplets of random image patches.}
	\label{fig:result}
\end{figure}

The main simulations focused on developing a denoising model using the proposed hybrid framework and classical Pix2Pix as the baseline to produce high-quality results. The two models showed comparable performances in terms of denoising. Then, to test the robustness against mode collapse, a further re-training phase with different parameters was executed to try to induce mode collapse. The output denoised patches from the best-performing model demonstrate the capabilities of this methodology, as shown in Fig. \ref{fig:result}.

The smoother training curve for discriminator loss leads to a more controlled learning process, which is produced by smooth gradient updates in the discriminator’s loss. This allows the generators to exhibit smoother behavior. For validation, the Structural Similarity Index (SSIM) was computed between the artificially denoised and original clean samples, as shown in the expression below. The Structural Similarity Index (SSIM) between a clean image $x = x_{\text{clean}}$ and a denoised image $y = x_{\text{denoised}}$ is given by:

\begin{equation}
	\text{SSIM}(x, y) = \frac{(2\mu_x\mu_y + c_1)(2\sigma_{xy} + c_2)}{(\mu_x^2 + \mu_y^2 + c_1)(\sigma_x^2 + \sigma_y^2 + c_2)}
\end{equation}
Where:
\begin{itemize}
	\item \( \mu_x \) and \( \mu_y \) are the averages of \( x \) and \( y \) respectively.
	\item \( \sigma_x^2 \) and \( \sigma_y^2 \) are the variances of \( x \) and \( y \) respectively.
	\item \( \sigma_{xy} \) is the covariance of \( x \) and \( y \).
	\item \( c_1 \) and \( c_2 \) are constants to avoid division by zero, often defined as \( c_1 = (k_1L)^2 \) and \( c_2 = (k_2L)^2 \), with \( L \) being the dynamic range of the pixel values (commonly \( L = 255 \) for 8-bit images), and \( k_1 = 0.01 \) and \( k_2 = 0.03 \).
\end{itemize}

Other metrics introduced to measure noise reduction alongside SSIM include Peak Signal-to-Noise Ratio (PSNR) and Mean Squared Error (MSE). These measures provide additional insights into the effectiveness of noise reduction techniques. However, PSNR can sometimes offer limited information about local details in images. To avoid undesirable outcomes, such as the loss of essential text elements during denoising, SSIM is still employed due to its ability to maintain structural information in image data. With robust hyper-parameter tuning, the hybrid model achieves an SSIM score of over 95\% on testing data. Additionally, PSNR and MSE have been incorporated to provide a more comprehensive evaluation of the models. To validate the proposed model and compare it with a baseline, SSIM, PSNR, and MSE were assessed using both the proposed hybrid WGAN-GP Pix2Pix and the classical Pix2Pix. The SSIM, PSNR, and MSE values were calculated and averaged across a sample of denoised and ground-truth clean images for both methods. The results are summarized in Table \ref{tab2}:

\begin{table}[hbt!]
	\begin{center}
		\caption{Averaged SSIM Scores on Testing Dataset \label{tab2}}
		\begin{tabular}{| c | c | c |} 
			\hline
			& WGAN-GP Pix2Pix (ours) & Classical Pix2Pix \\ [0.5ex] 
			\hline 
			SSIM & \textbf{0.9581 \% }& 0.9416 \% \\ 
			\hline
			PSNR & \textbf{20.90 \% }& 19.06 \% \\ 
			\hline
			MSE & \textbf{775.14 \% }& 1122.64 \% \\ 
			\hline
		\end{tabular}
	\end{center}
	\vspace{-4mm}
\end{table}

\begin{figure*}[!htb]
	\centering
	\centering
    \includegraphics[scale=0.30]{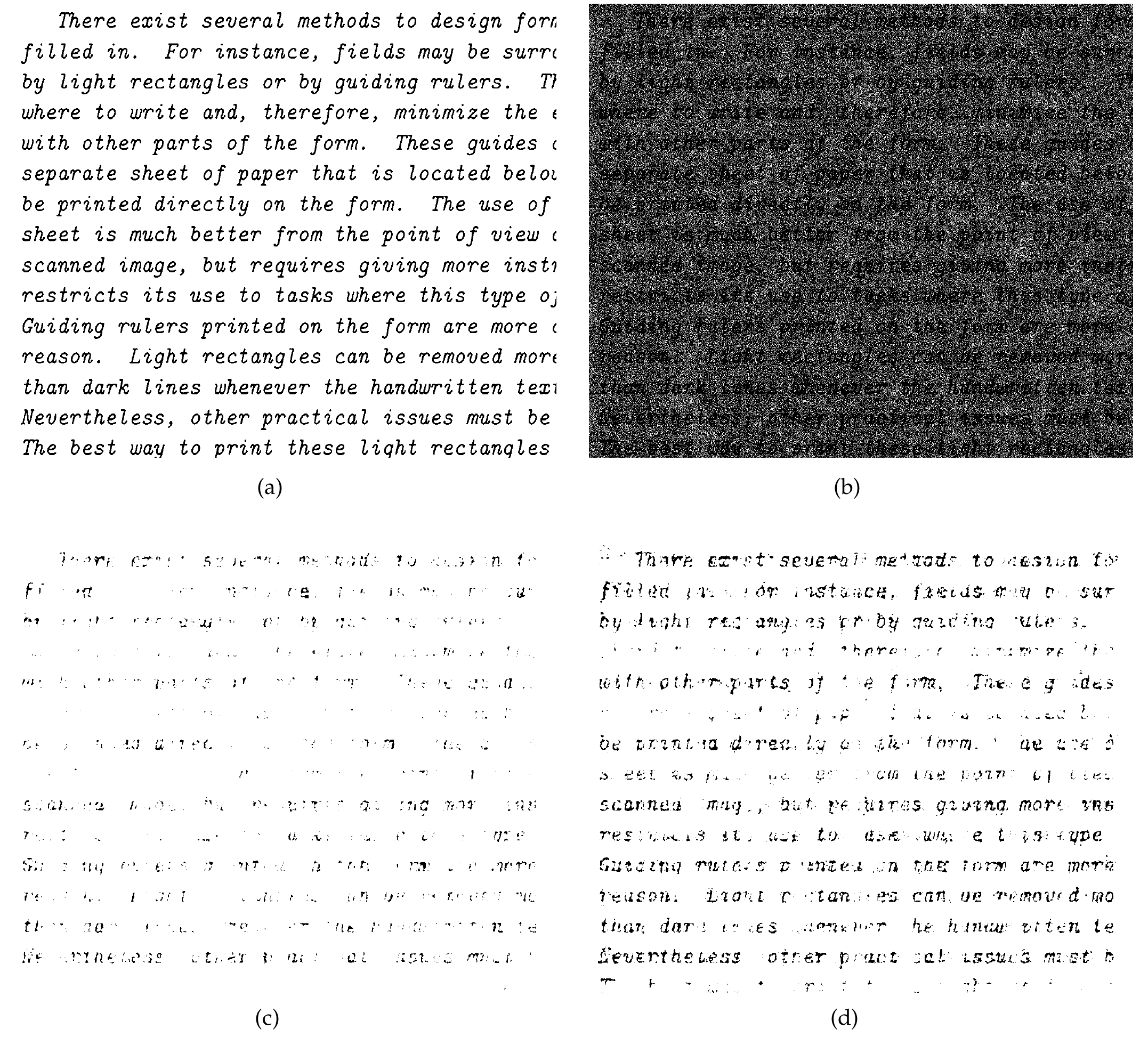}
	\caption{Parts (a) and (b) show instances of the original clean document and its noisy counterpart from the dataset, respectively. Part (c) illustrates the denoising performance of the same instance by the classical Pix2Pix model, while Part (d) depicts the output of the proposed WGAN-GP Pix2Pix.}
	\label{fig:Result1}
\end{figure*}

To provide a visual representation of the denoising process, we have included a comparison of the output of the two models on the same testing image, as shown in figure \ref{fig:Result1}.

\begin{figure}[]
	\centering
	\includegraphics[scale=0.29]{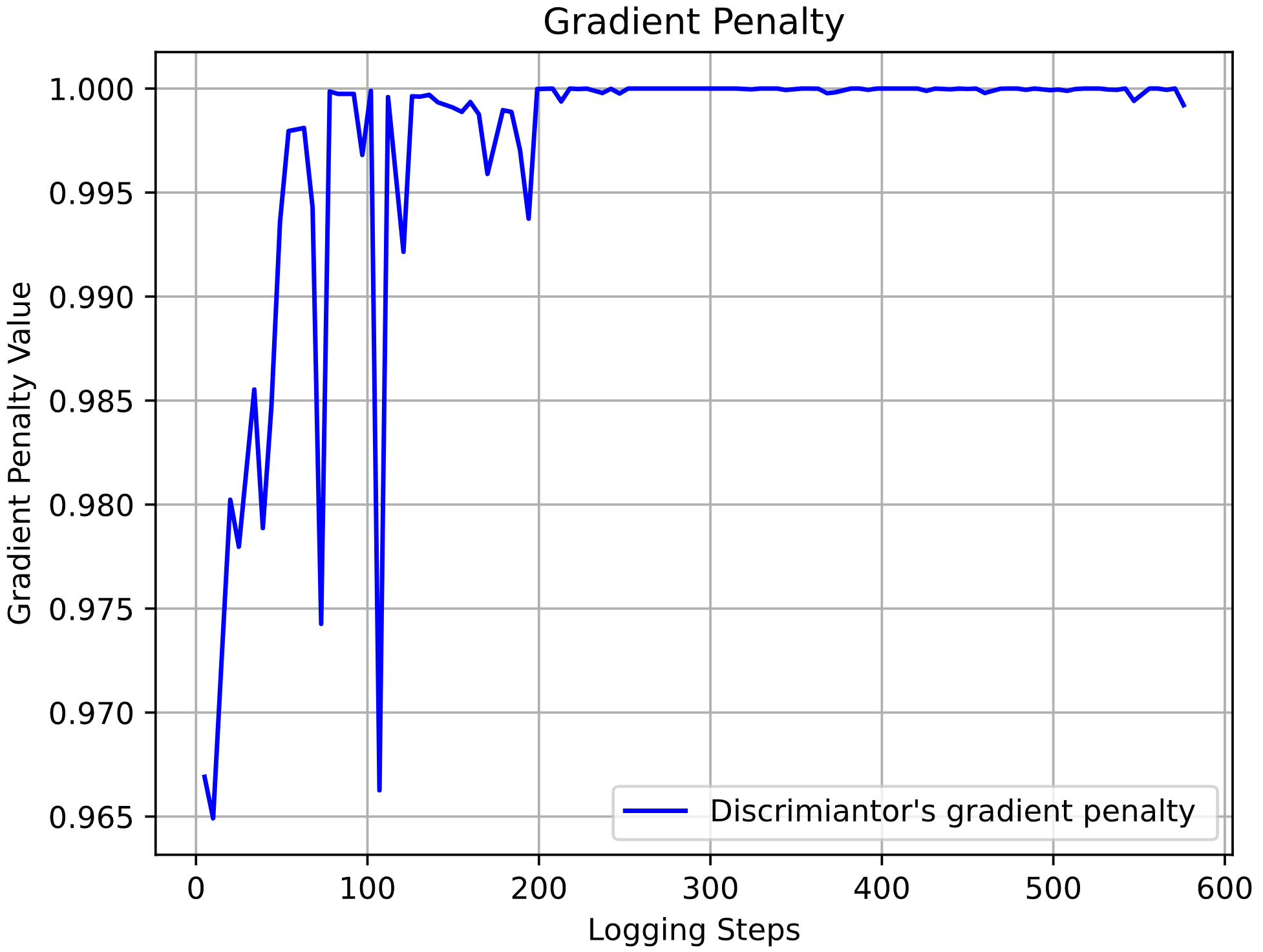}
	\caption{Discriminator's Loss Gradient Penalty Component.}
	\label{fig:gradpen}
\end{figure}

\begin{figure}[]
	\centering
	\includegraphics[scale=0.29]{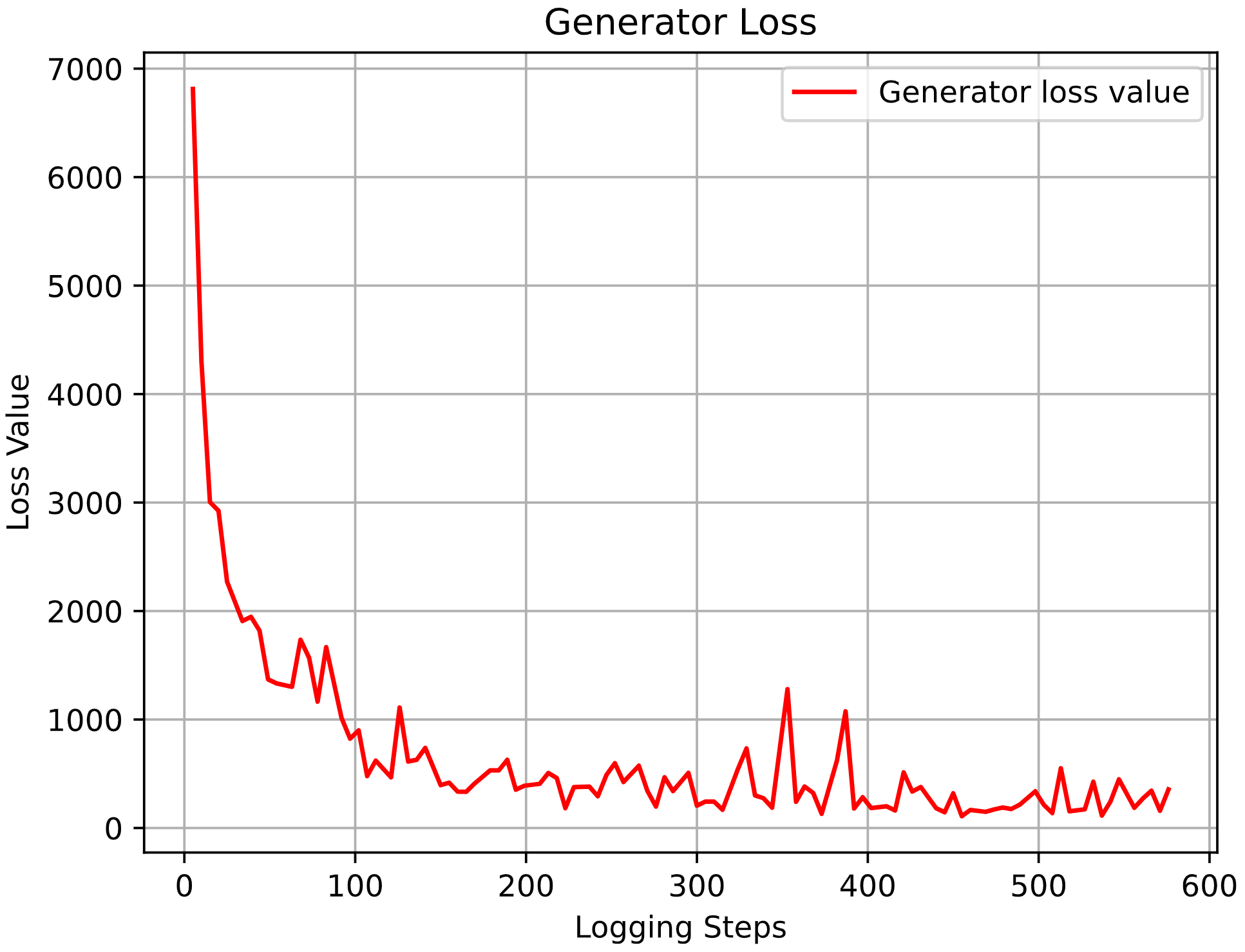}
	\caption{Generator Network Training Loss using Pix2Pix WGAN-GP.}
	\label{fig:genloss}
\end{figure}

\begin{figure}[]
	\centering
	\includegraphics[scale=0.29]{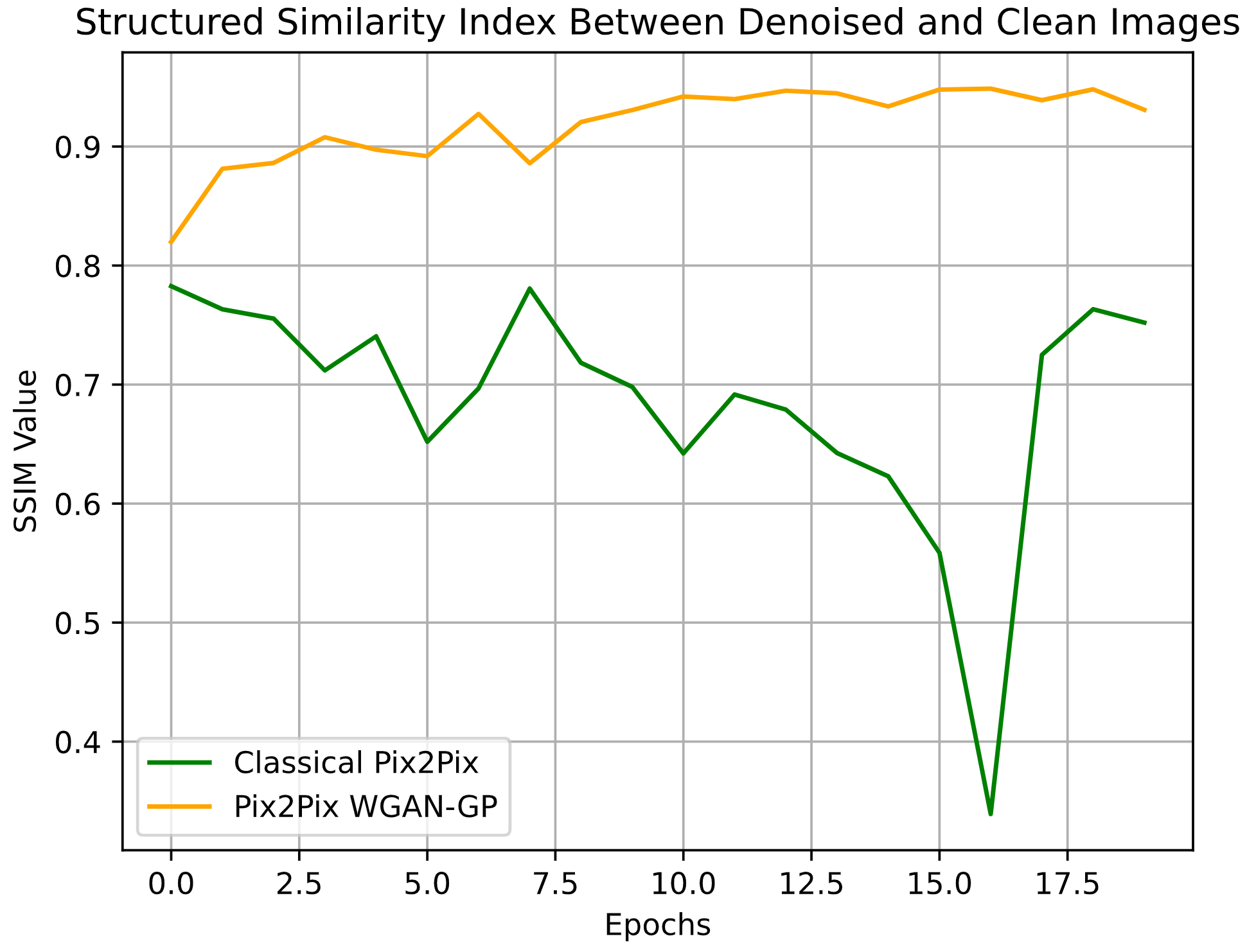}
	\caption{Structured Similarity Index Values during Inference Steps on Validation Data during a Mode Collapse.}
	\label{fig:ssim}
\end{figure}

\begin{figure}[]
	\centering
	\includegraphics[scale=0.29]{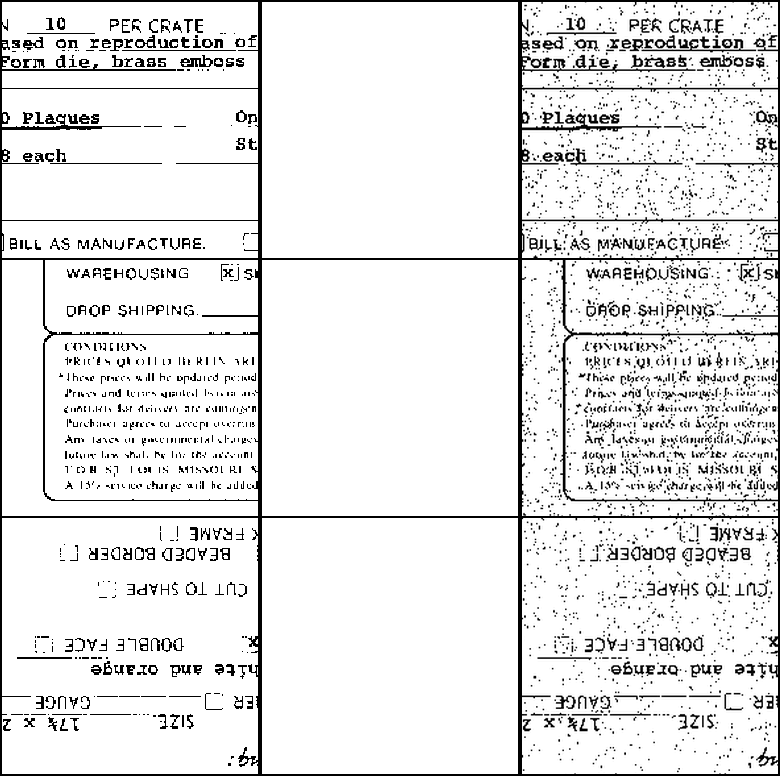}
	\caption{An Example of clean-denoised-noisy image patches experiencing Mode Collapse in GANs' Training.}
	\label{fig:collapse_mode}
\end{figure}

In Fig. \ref{fig:gradpen}, the gradient penalty component during learning is illustrated. It can be deduced that during the learning process, the discriminator outperformed the generator, causing the gradient penalty to saturate at one. This was particularly evident toward the end of the training phase, slowing down the discriminator's learning process by enforcing the Lipschitz constraint. In Fig. \ref{fig:genloss}, the generator loss of the WGAN Pix2Pix hybrid is presented, showing convergence with a smoother curve compared to the classical Pix2Pix loss. Given that the objective of this research was to highlight the susceptibility of classical GANs to mode collapse and demonstrate how this hybrid version mitigates it, both models were subsequently retrained with altered parameters to intentionally induce mode collapse. Fig. \ref{fig:ssim} was obtained by performing validation steps after each epoch and computing SSIM between denoised and noisy images. This illustrates the accuracy drop during classical Pix2Pix model training. To substantiate this challenge, during the SSIM drop, the model's outputs were inspected. The model consistently generated white patches, regardless of the input it received to be denoised, as shown in Fig. \ref{fig:collapse_mode}.

Moreover, it can be observed that mode collapse affect the training from the early stages, producing a slowly decreasing SSIM accuracy, difficult to detect. Interestingly, towards the end of the learning process, the model appeared to spontaneously recover from mode collapse. However, such behavior is relatively uncommon, and typically, after such occurrences, training often stagnates and fails to achieve convergence. All the code developed for this work can be found at \url{https://github.com/lucatirel/pix2pix-wgangp}.

\section{Conclusions \& Future Work \label{sec:Conclusions}}
In this paper, a hybrid model that combines classical Pix2Pix elements with Wasserstein loss and gradient penalty was explored. The primary objectives were to counteract the stability issues prevalent during GAN training and to reduce reliance on exhaustive hyperparameter tuning. The proposed hybrid method seamlessly integrates the Pix2Pix architecture with the Wasserstein loss and gradient penalty. The superiority of the GAN approach is evident from the denoising results, where accuracy in validation reaches above 94\% for both models under an optimal training regimen. A salient feature of the WGAN hybrid version is its resilience against mode collapse, a recurrent obstacle in GAN training that can compromise stability. The hybrid methodology promotes a smoother learning trajectory due to the discriminator’s loss, fortifying learning dynamics and substantially minimizing the risk of insidious mode collapses.
	While classical Pix2Pix excels at contextually aligned image generation but may suffer from instability with complex datasets, WGAN-GP provides stable training and reduces mode collapse. The integrated model benefits from Pix2Pix’s feature mapping and WGAN-GP’s robust loss function, resulting in higher fidelity denoised images and better generalization across various noise conditions.\\

	This investigation focused on the denoising of binary images, which inherently possess lower information content than their grayscale counterparts, given their pixel values are restricted to 0 or 1. This limitation, along with the computational costs involved, rules out several advanced methodologies such as Connected Components Analysis (CCA) combined with shape information to filter noise, as well as frequency domain analysis and disk-based filtering. Furthermore, it necessitates leveraging neural network non-linearities to discern and learn intricate nonlinear patterns in datasets that are not characterized by linear separability. A quintessential challenge is discerning nuanced features, such as preserving dots atop the letter 'i' while eliminating other dot-like noise—a task aptly suited for neural networks. Both proposed approaches are adept at processing binary images. Moreover, noise can be synthetically layered on pristine images to cultivate artificial datasets. Future research will venture into the realm of Matching Moments Networks, aiming to further enhance the stability properties of the discriminator and explore the potentials of generative transformers.

	\subsection{Future Works}
	The proposed hybrid architecture was trained on several randomly generated noise types, varying in their intensities, standard deviations, locations on the document, and noise types. One of the next steps for this work will be to construct a wider dataset with a larger set of noise types to help the model generalize even better. Moreover, the authors will focus on making the binary denoising even more robust to mode collapse by adopting and comparing other types of networks, such as generative moment matching networks.

\section*{Acknowledgments}

This work was supported in part by the National Sciences and Engineering Research Council of Canada (NSERC), under the grants RGPIN-2022-04937.

\section*{Credit authorship contribution statement}
Luca Tirel: Writing the original draft, Investigation, Methodology, Conceptualization, Software, Methodology, Visualization, Formal
analysis.
Ali Mohamed Ali: Writing the original draft, Investigation, Methodology, Software, Methodology, Visualization, Formal
analysis.
Hashim A. Hashim: Writing review \& editing, Conceptualization, Visualization, Supervision, Investigation,  Validation, Funding acquisition.

\section*{Data availability}
Data will be made available on request.

\section*{Declaration of competing interest}
The authors declare that they have no known competing financial interests or personal relationships that could have appeared to influence the work reported in this paper.

\balance
\bibliographystyle{IEEEtran}
\bibliography{ref}
		
	\end{document}